\documentclass[showpacs,twocolumn,superscriptaddress,pra]{revtex4-1}
\usepackage[T1]{fontenc}
\usepackage[utf8]{inputenc}
\usepackage{amsmath}
\usepackage{amssymb}
\usepackage{dsfont}
\usepackage{color}
\makeatletter
\usepackage{graphicx}
\usepackage{enumerate}
\graphicspath{{./figures/}}
\makeatother

\newcommand{\mri}{\mathrm{i}}
\newcommand{\Exp}[1]{\mathrm{e}^{\mbox{\footnotesize$#1$}}}

\renewcommand{\vec}[1]{\mathbf{#1}}

\begin{document}
\title{Phase Separation and Dynamics of two-component Bose-Einstein condensates}

\author{Kean Loon Lee}
\affiliation{Joint Quantum Centre (JQC) Durham-Newcastle, School of Mathematics and Statistics,
Newcastle University, Newcastle upon Tyne NE1 7RU, England, UK}

\author{Nils B. J\o rgensen}
\affiliation{Institut for Fysik og Astronomi, Aarhus Universitet, Ny Munkegade 120, DK-8000 Aarhus C, Denmark}

\author{I-Kang Liu}
\affiliation{Joint Quantum Centre (JQC) Durham-Newcastle, School of Mathematics and Statistics,
Newcastle University, Newcastle upon Tyne NE1 7RU, England, UK}
\affiliation{Department of Physics, National Changhua University of Education,
Changhua 50058, Taiwan}

\author{Lars Wacker}
\author{Jan J. Arlt}
\affiliation{Institut for Fysik og Astronomi, Aarhus Universitet, Ny Munkegade 120, DK-8000 Aarhus C, Denmark}

\author{Nick P. Proukakis}
\affiliation{Joint Quantum Centre (JQC) Durham-Newcastle, School of Mathematics and Statistics,
Newcastle University, Newcastle upon Tyne NE1 7RU, England, UK}

\pacs{}
\date{\today{}}

\begin{abstract}\noindent
The miscibility of two interacting quantum systems is an important testing ground for the understanding of complex quantum systems. Two-component Bose-Einstein condensates enable the investigation of this scenario in a particularly well controlled setting. In a homogeneous system, the transition between mixed and separated phases is fully characterised by a ‘miscibility parameter’, based on the ratio of intra- to inter-species interaction strengths. Here we show,  however, that this parameter is no longer the optimal one for trapped gases, for which the location of the phase boundary depends critically
on atom numbers. We demonstrate how monitoring of damping rates and frequencies of dipole oscillations enables the experimental mapping of the phase diagram by numerical implementation of a fully self-consistent finite-temperature kinetic theory for binary condensates. The change in damping rate is explained in terms of surface oscillation in the immiscible regime, and counterflow instability in the miscible regime, with collisions becoming only important in the long time evolution.

\end{abstract}
\pacs{03.75.Mn,67.85.-d,03.75.Kk}
\maketitle

\emph{Introduction.}
Bose-Einstein condensates (BECs) are attractive systems to study the non-equilibrium dynamics of interacting quantum gases~\cite{weiner_bagnato_1999,*dalfovo_giorgini_1999,*bloch_dalibard_2008}. The large variety of atomic BECs that have been produced experimentally, and the ability to engineer tailored potentials~\cite{grimm_weidemuller_2000,gaunt_schmidutz_2013} as well as to control interactions~\cite{chin_grimm_2010}, offer a plethora of research directions. A fascinating possibility is the creation of binary mixtures using different hyperfine levels ($^{87}$Rb~\cite{myatt_burt_1997,hall_matthews_1998,maddaloni_modugno_2000}), different isotopes (e.g. $^{87}$Rb-$^{85}$Rb~\cite{papp_pino_2008}, $^{168}$Yb-$^{174}$Yb~\cite{sugawa_yamazaki_2011}) , different elements (e.g.~$^{87}$Rb-$^{41}$K~\cite{modugno_modugno_2002,*thalhammer_barontini_2008}, $^{87}$Rb-$^{133}$Cs~\cite{mccarron_cho_2011,lercher_takekoshi_2011}, $^{87}$Rb-$^{84}$Sr and $^{87}$Rb-$^{88}$Sr~\cite{pasquiou_bayerle_2013}, $^{87}$Rb-$^{39}$K~\cite{wacker_jorgensen_2015,minardi_private}, $^{87}$Rb-$^{23}$Na~\cite{wang_li_2016}) or even atoms with different statistics (e.g. $^{6}$Li-$^{7}$Li~\cite{ferrier-barbut_delehaye_2014}). Experimentally, these mixtures have been used to explore such diverse issues as collective modes~\cite{maddaloni_modugno_2000,ferrier-barbut_delehaye_2014}, pattern formation~\cite{hamner_chang_2011,*hoefer_chang_2011,de_campbell_2014}, phase separation~\cite{papp_pino_2008,mccarron_cho_2011,wacker_jorgensen_2015,wang_li_2016}, nonlinear dynamical excitations~\cite{mertes_merrill_2007,eto_takahashi_2015a,*eto_takahashi_2015b}, Kibble-Zurek mechanism~\cite{nicklas_karl_2015} and the production of dipolar molecules~\cite{molony_gregory_2014}.

\begin{figure}[!b]
  \includegraphics[width=0.49\textwidth]{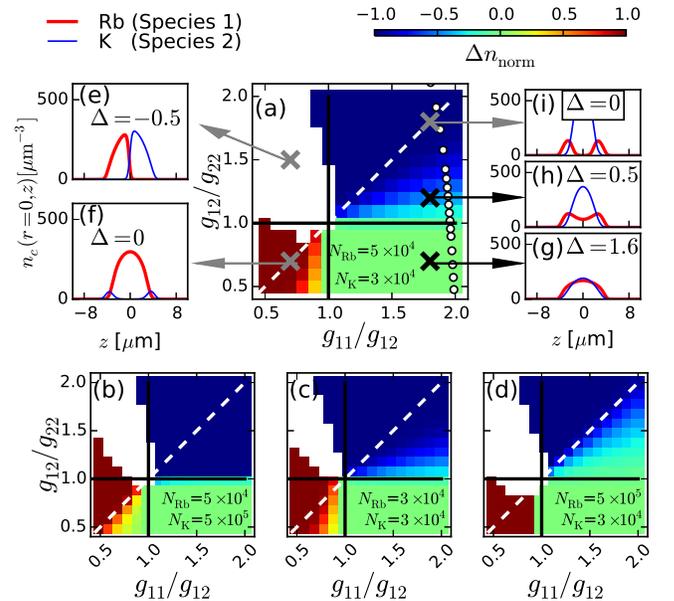}
   \vspace*{-5mm}
  \caption{\label{fig:aniso_phase} (Color online) (a)--(d) Ground state phase diagram of a trapped $^{87}$Rb-$^{39}$K mixture at temperature $T=0$ for various total numbers of $^{87}$Rb ($N_{\rm Rb}$, species 1) and $^{39}$K ($N_{\rm K}$, species 2) atoms. The symmetrical phase is characterised by a normalised trap-center density $\Delta n_{\rm norm}$ [see Eq.~\eqref{eq:n_norm}] while the asymmetrical phase is shown in white. Typical density profiles along the $z$-axis are shown in (e) for asymmetrical phase and (f)--(i) for symmetrical phase. White circles in (a) indicates points accessible experimentally using the Feshbach resonances reported in Ref.~\cite{wacker_jorgensen_2015}.}
\end{figure}

An important property of a binary mixture is its miscibility. For a homogeneous system, the miscible-immiscible transition is uniquely characterised by the miscibility parameter $\Delta=(g_{11}g_{22}/g_{12}^2) - 1$. Depending on the strength of the intra- ($g_{11}$, $g_{22}$) and inter-species ($g_{12}$) interaction, the two components can either overlap in space ($\Delta>0$) or phase separate ($\Delta<0$). This spatial overlap has practical consequences on, e.g. rethermalisation rate~\cite{delannoy_murdoch_2001}, coarse-graining dynamics~\cite{ronen_bohn_2008,hofmann_natu_2014,pattinson_parker_2014,*liu_pattinson_2016}, structures of vortex lattice~\cite{mason_aftalion_2011,*aftalion_mason_2012} or instabilities in fluid dynamics~\cite{kasamatsu_tsubota_2004,*takeuchi_ishino_2010,*suzuki_takeuchi_2010}.
Based on the assumption of overlapping trap centers of the two components, numerical studies~\cite{ho_shenoy_1996,pu_bigelow_1998,ohberg_1999,trippenbach_goral_2000,pattinson_billam_2013,polo_ahufinger_2015}  have shown three different types of density profiles [Fig.~\ref{fig:aniso_phase}(e)--(i)]: a mixed phase where both components overlap at the trap center; a symmetrical demixed phase where one component forms a shell structure around the other; and an asymmetrical demixed phase where the centers-of-mass (COMs) of the two components do not coincide. Although such features were also seen experimentally \cite{tojo_taguchi_2010,mccarron_cho_2011}, a full characterisation of the boundaries separating the three phases is still missing.

The aim of this article is fourfold: 
(i) to demonstrate that, for a trapped binary mixture, $\Delta=0$ 
is generally no longer the optimal criterion for the transition boundary;
(ii) to characterise the full phase diagram [see Fig.~\ref{fig:aniso_phase}] based on the identification of a new parameter;
(iii) to propose measurements of the frequency and damping rates of induced dipole oscillations as a universal experimental tool for mapping out the phase diagram; and
(iv) to demonstrate the importance of thermal effects on the dynamics, by providing the first numerical impementation of a fully self-consistent finite-temperature model for binary mixtures \cite{edmonds_lee_2015a,edmonds_lee_2015b}. This extends the successful `Zaremba-Nikuni-Griffin' (ZNG) model~\cite{zaremba_nikuni_1999,*nikuni_zaremba_1999,*griffin_nikuni_2009} to two components. The validity of the ZNG model has previously been verified in studies of collective modes of single-component BECs~\cite{jackson_zaremba_2001,*jackson_zaremba_2002a,jackson_zaremba_2002} and macroscopic excitations~\cite{jackson_proukakis_2007,*jackson_barenghi_2007,jackson_proukakis_2009,*allen_zaremba_2013}.

Typical studies to date have probed the phases by varying a single parameter (e.g. $g_{12}$~\cite{tojo_taguchi_2010,ohberg_1999} or the number of atoms~\cite{mccarron_cho_2011,pu_bigelow_1998}), equivalent to a line scan in a multi-dimensional parameter space. In these investigations it was thus impossible to obtain a full picture of the problem. Here we present a ground state phase diagram of the density profiles in a four-dimensional space, treating the ratios of the interaction strengths ($g_{11}/g_{12}$, $g_{12}/g_{22}$) and the numbers of atoms ($N_{\rm Rb}$, $N_{\rm K}$) as independent variables [Fig.~\ref{fig:aniso_phase}(a)--(d)]. 
Not only do we find
a clear deviation of the boundaries from the uniform case ($g_{11}/g_{12}=g_{12}/g_{22}$), which depends on the atom numbers, but we can also predict whether certain phases (e.g. the asymmetrically-separated phase) are accessible to a particular mixture or experiment.

The mapping of the transition boundary in experiments is a non-trivial task. A typical procedure~\cite{papp_pino_2008,wacker_jorgensen_2015,wang_li_2016} involves free expansion of the BECs and a measurement of the separation of the COMs of the expanded clouds as a function of $\Delta$. The boundary is then identified as the point where the separation starts to grow. While this method yields quantitative agreement with the transition boundary in the uniform case, we have previously~\cite{wacker_jorgensen_2015} shown that the measured separation can be influenced by the repulsion developed during the expansion dynamics rather than the in-trap phase separation. In the second part of this work, we therefore propose to map the transition boundary of a trapped mixture by monitoring the damping rate and the frequency of dipole oscillations, which we numerically find to be sensitive to the in-trap density profiles. Based on experimentally relevant parameters, our numerical simulations of the oscillation dynamics at both zero and finite temperature indicate an abrupt increase in both the damping rate and the frequency when crossing from the immiscible to the miscible regime, which generally occurs at $\Delta \neq 0$.

\emph{Model.}
To address the role of temperature in such dynamics, we use the 2-component generalisation of the ZNG model, which has been previously demonstrated to 
pass the stringent test of the undamped Kohn mode, essential for a correct modelling of collective modes~\cite{jackson_zaremba_2002b}. 
Our kinetic model \cite{edmonds_lee_2015a,*edmonds_lee_2015b} describes the self-consistent coupling of two BECs, each coupled to their own thermal cloud, and additionally includes coupling between the thermal clouds. This approach enables us to consider the relative importance of damping arising from mean-field coupling ($U_c^j$, $U_n^j$) and thermal-condensate ($C_{12}^{..}$, $\mathds{C}^{kj}_{12}$) or thermal-thermal ($C_{22}^{..}$) collisions. 
Each condensate wavefunction, $\phi_j(\vec{r})$ obeys a dissipative Gross-Pitaevskii equation
\begin{equation}
i\hbar\frac{\partial\phi_j}{\partial t}=\bigg[-\frac{\hbar^2}{2m_j}\nabla^2+U^{j}_{c}-i(R^{jj}+R^{kj}+\mathds{R}^{kj})\bigg]\phi_j,\label{eq:dschro1}
\end{equation}
while the Wigner distribution function of the thermal atoms $f^j(\vec{p},\vec{r},t)$ obeys a quantum Boltzmann equation
\begin{align}\nonumber\label{eq:qbe}
&\frac{\partial}{\partial t}f^{j}+\frac{1}{m_j}{\bf p}\cdot\nabla_{\bf r}f^{j}-\nabla_{\bf p}f^{j}\cdot\nabla_{\bf r}U^{j}_{\text{n}}\\&=\bigg(C^{jj}_{12}+C^{kj}_{12}\bigg)+\mathds{C}^{kj}_{12}+\bigg(C^{jj}_{22}+C^{kj}_{22}\bigg).
\end{align}
The effective potentials are related to the condensate density $n_{c,j}(\vec{r}) = |\phi_j(\vec{r})|^2$, and thermal atom density $\tilde{n}_j(\vec{r}) = \int d\vec{p}/(2\pi\hbar)^3 f^j(\vec{p},\vec{r},t)$, as
\begin{subequations}
\begin{align}
U_c^j =& V_j+g_{jj}(n_{c,j} +2\tilde{n}_j)+g_{kj}(n_{c,k} +\tilde{n}_k),\\
U_n^j =& V_j+ 2g_{jj}(n_{c,j} + \tilde{n}_j) +g_{kj}(n_{c,k} + \tilde{n}_k).
\end{align}
\end{subequations}
Here $g_{kj}=2\pi\hbar^2 a_{kj}/m_{kj}$ denotes the effective interaction strength, where $a_{kj}$ defines the s-wave scattering length between atoms in components $j$ and $k$, and $m^{-1}_{kj}=m^{-1}_{j}+m^{-1}_{k}$ defines the reduced mass. 
The source terms in Eq.~\eqref{eq:dschro1} are related to the first 3 collision integrals in Eq.~\eqref{eq:qbe} via $R^{kj}({\bf r},t)=\frac{\hbar}{2n_{c,j}}\int\frac{d{\bf p}}{(2\pi\hbar)^3}C^{kj}_{12}$ ($k=j$, $k\neq j$) and $\mathds{R}^{kj}({\bf r},t)=\frac{\hbar}{2n_{c,j}}\int\frac{d{\bf p}}{(2\pi\hbar)^3}\mathds{C}^{kj}_{12}$. These collision integrals are sampled in dynamical simulation through the direct simulation Monte Carlo~\cite{jackson_zaremba_2002} method.

\emph{Simulation parameters.} For our simulations, we consider the $^{87}$Rb-$^{39}$K mixture~\cite{wacker_jorgensen_2015}, but our analysis is general and can be extended to other mixtures. 
The atoms are confined in a harmonic potential $V_j(\vec{r})=\frac{1}{2}m_j[\omega_{r,j}^2 r^2 + \omega_{z,j}^2 (z-z_j)^2]$ with radial and axial angular frequencies, $\omega_{r,j}=2\pi \nu_{r,j}$ and $\omega_{z,j}=2\pi\nu_{z,j}$, respectively, with 
$\nu_r=119\,$Hz (178$\,$Hz) and $\nu_z=166\,$Hz (248$\,$Hz) for $^{87}$Rb ($^{39}$K) atoms -- labelled as species 1 (2) -- in such a way that $m_1\omega_{r,1}^2 = m_2\omega_{r,2}^2$ (and similarly for the axial frequencies). For atomic species with different masses, there is a gravitational sag between the two trap centers, i.e. $|z_1 - z_2| > 0$. Since this sag can be eliminated, we treat the sag $z_1-z_2$ as a variable in our modelling. The scattering length of $^{87}$Rb is fixed at $a_{11}=99\,a_0$.


\emph{Equilibrium phase diagram.} At equilibrium, both the source terms and the collision integrals vanish, hence we can set $\phi_j(\vec{r},t) = \phi_j(\vec{r})\Exp{-\mri \mu_j t/\hbar}$ with chemical potential $\mu_j$ in Eq.~\eqref{eq:dschro1} and use a semi-classical Hartree-Fock approximation for the thermal cloud 
$\tilde{n}_j(\vec{r}) = g_{3/2}(\Phi_j)/\lambda_j^3$ with thermal wavelength $\lambda_j = \sqrt{2\pi\hbar^2/(m_j k_B T)}$ and local fugacity $\Phi_j = \exp[(\mu_j-U_n^j)/(k_B T)]$ at temperature $T$. We also set the sag to be zero. We then obtain the density profiles by solving Eq.~\eqref{eq:dschro1} (using imaginary-time propagation) and $\tilde{n}_j(\vec{r})$ self-consistently.

\begin{figure}[!t]
  \includegraphics[width=0.49\textwidth]{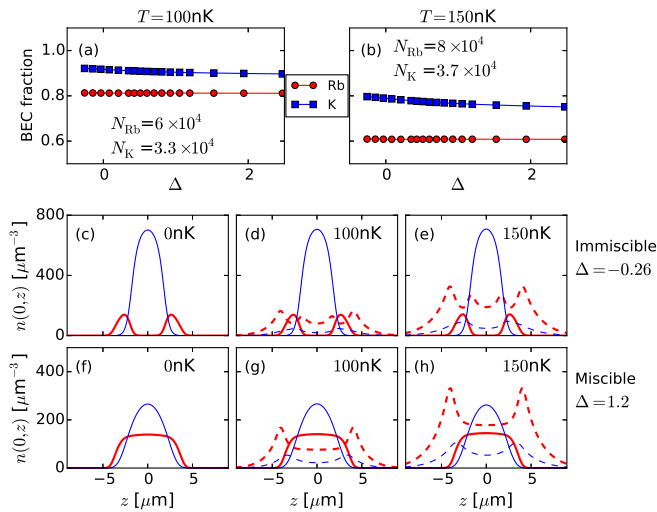}
    \vspace*{-5mm}
  \caption{\label{fig:aniso_frac_den} (Color online) Results with approximately $5\times10^4$ $^{87}$Rb and $3\times10^4$ $^{39}$K BEC atoms. (top) Condensate fractions of a mixture with indicated total atom numbers at (a) $T=100\,$nK (non-interacting critical temperatures $T_{c,1} = 234\,$nK and $T_{c,2}= 287\,$nK) and (b) $T=150\,$nK ($T_{c,1}= 259\,$nK and $T_{c,2} = 298\,$nK), as a function of miscibility parameter $\Delta$. (middle) Axial densities of an immiscible mixture of $^{87}$Rb (thick red) and $^{39}$K (thin blue) atoms at different temperatures. Solid (dashed) line represents condensate (thermal cloud). The thermal cloud densities have been amplified by 20 times for clarity. (bottom) Same as middle panels but for a miscible mixture.}
\end{figure}

Focussing initially on the zero-temperature case, we consider four different values ($3\times10^4, 5\times10^4,10^5, 5\times10^5$) for the total number of $^{87}$Rb atoms ($N_{\rm Rb}$) and $^{39}$K atoms ($N_{\rm K}$), resulting in a total of sixteen combinations. 
The ground states, shown in Fig.~\ref{fig:aniso_phase}(e)--(i), are then given by the final density profiles with the lowest total energy, probed numerically by means of different initial states~\footnote{In order to probe the possible ground state with asymmetrical density distribution, we perform two sets of simulations: one starts with wave functions as single-component Thomas-Fermi profiles centered at $z=0$, while the other starts with Gaussian wave functions of width $3\mu$m and separated along the $z$-axis by 6$\mu$m.}.


The ambiguity in classifying a trapped mixture into a mixed or demixed phase becomes evident in Fig.~\ref{fig:aniso_phase}(h). Based on the usual prescription (motivated by the homogeneous system), this would typically be classified as a mixed phase, since $\Delta=0.5$. To avoid such an evident inconsistency, we instead propose characterising the mixture (with symmetrical distribution) by the difference in normalised trap-center density, 
\begin{align}
  \Delta n_{\rm norm} = \frac{n_{c,1}(\vec{0})}{\max n_{c,1}(\vec{r})} - \frac{n_{c,2}(\vec{0})}{\max n_{c,2}(\vec{r})}.\label{eq:n_norm}
\end{align}
If the two components overlap strongly [e.g. Fig.~\ref{fig:aniso_phase}(g)], we have $\Delta n_{\rm norm}=0$. On the other hand, if the two components repel each other so strongly that one acts as an impenetrable barrier to the other [e.g. Fig.~\ref{fig:aniso_phase}(f),~\ref{fig:aniso_phase}(i)], we have $\Delta n_{\rm norm}=1(-1)$ if $^{87}$Rb ($^{39}$K) sits at the trap center. For ground states with asymmetrical distribution [e.g. Fig.~\ref{fig:aniso_phase}(e)], we have non-zero separation, i.e. $\langle z\rangle_1\neq\langle z\rangle_2$.

The $g_{12}/g_{22}$--$g_{11}/g_{12}$ space is thus divided into four regions, as marked by the solid black lines in Fig.~\ref{fig:aniso_phase}(a)-(d), depicting the phase diagram for $N_{\rm Rb}=5\times10^4$ and various $N_{\rm K}$. These are: a miscible phase (bottom right), a strongly-immiscible phase with asymmetrical distribution (top left), and two regions that contain symmetrical distributions with shell-structure (bottom left and top right). It is interesting to notice how the boundary between $|\Delta n_{\rm norm}|=1$ and $|\Delta n_{\rm norm}|<1$ deviates from the homogeneous criterion $\Delta=0$ (dashed white lines). In particular, as $N_{\rm K}$ increases [Fig~\ref{fig:aniso_phase}(a)$\to$(b)], the boundary rotates clockwise from $\Delta=0$ into $g_{12}=g_{22}$, which coincides with the black horizontal line. Likewise, if we fix $N_{\rm K}$ but increase $N_{\rm Rb}$ [Fig~\ref{fig:aniso_phase}(c)$\to$(a)$\to$(d)], the boundary rotates anti-clockwise. This observation can be understood by considering the $N_{\rm K}\gg N_{\rm Rb}$ limit. To the lowest order we neglect the effect of $^{87}$Rb on $^{39}$K, hence we can adopt the single-component Thomas-Fermi approximation, $n_{c,2}\approx [\mu_2 - \frac{m_2}{2} (\omega_{r,2}^2 r^2 + \omega_{z,2}^2 z^2)]/g_{22}$. Taking into account the mean field $g_{12}n_{c,2}$, $^{87}$Rb atoms experience an effective trap potential $ (1-g_{12}/g_{22}) m_1(\omega_{r,1}^2 r^2 + \omega_{z,1}^2 z^2)/2$ at the trap center. If $g_{12}/g_{22}<1$, the effective trap remains harmonic, hence $^{87}$Rb has a peak at the center. However, if $g_{12}/g_{22}>1$, the effective trap turns into a potential barrier and $^{87}$Rb develops a shell structure. A complete phase diagram showing data for all sixteen combinations of $N_{\rm Rb}$ and $N_{\rm K}$ can be found in the appendix (Appendix Fig.~\ref{fig:aniso_phaseT0}) to further support our reasoning. We also show there (Appendix Fig.~\ref{fig:iso_phaseT0}) a similar observation for a phase diagram of an isotropic trap which rules out trap anisotropy as the underlying origin of our observations.

\begin{figure}[!t]
  \includegraphics[width=0.4\textwidth]{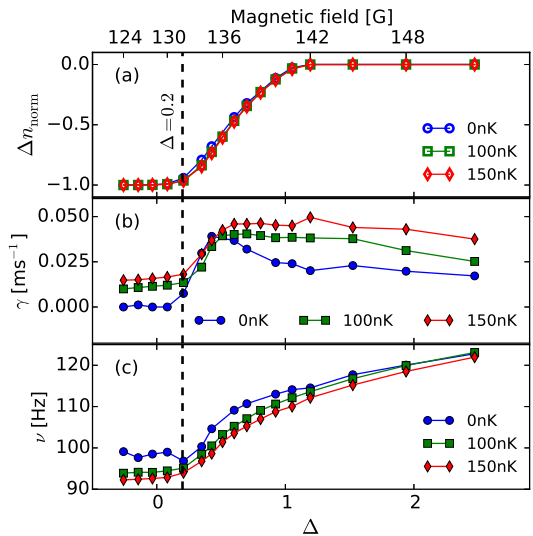}
  \vspace*{-5mm}
  \caption{\label{fig:aniso_den_damping} (Color online) (a) Difference in normalised trap-center density $\Delta n_{\rm norm}$, (b) damping rate $\gamma$ and (c) oscillation frequency $\nu$ of COM of $^{39}$K atoms as a function of miscibility $\Delta$ (bottom axis) or Feshbach magnetic field (top axis) at different temperature. The simulation parameters are the same as in Fig.~\ref{fig:aniso_frac_den}.}
\end{figure}

We next investigate how this boundary is changed at finite temperature. Consider a line-scan in the $g_{12}/g_{22}$--$g_{11}/g_{12}$ space that corresponds to tuning the scattering lengths $a_{12}$ and $a_{22}$ in the experimental setup~\cite{wacker_jorgensen_2015} by using Feshbach resonances. This is shown as the white circles in Fig.~\ref{fig:aniso_phase}(a). We simulate the density profiles at $T=100\,$nK $\leq0.4\,T_c$ and $150\,$nK $\leq0.6\,T_c$ for $N_{\rm Rb}=(6\textrm{--}8)\times10^4$,  and $N_{\rm K}=(3\textrm{--}4)\times10^4$, chosen such that the number of BEC atoms does not deviate significantly from $5\times10^4$ ($^{87}$Rb) and $3\times10^4$ ($^{39}$K) respectively [Fig.~\ref{fig:aniso_frac_den}(a)--(b)], used at $T=0\,$nK in Fig.~\ref{fig:aniso_phase}(a). We find the equilibrium condensate densities to be very similar for the three different temperatures for the same miscibility parameter [Fig.~\ref{fig:aniso_frac_den}(c)--(h)], even though the thermal clouds have different magnitudes. As a result, $\Delta n_{\rm norm}$ will be approximately the same even at different temperatures (Fig.~\ref{fig:aniso_den_damping}a) as long as the condensate atom numbers are comparable.

\emph{Detection through dipole oscillation.} The question arises whether $\Delta n_{\rm norm}$ is a physically-meaningful quantity. To investigate this, we dynamically simulate the dipole oscillation of the mixture at different miscibilities and temperatures. Starting with a mixture at equilibrium, we excite the dipole mode in our simulation by increasing the trap center separation from $0\,\mu$m to $0.2\,\mu$m in $5\,$ms linearly in time, followed by a decrease in the separation back to $0\,\mu$m in $1\,$ms linearly in time. The separation is chosen to be small compared to the Thomas-Fermi radii.

\begin{figure}[!b]
  \includegraphics[width=0.48\textwidth]{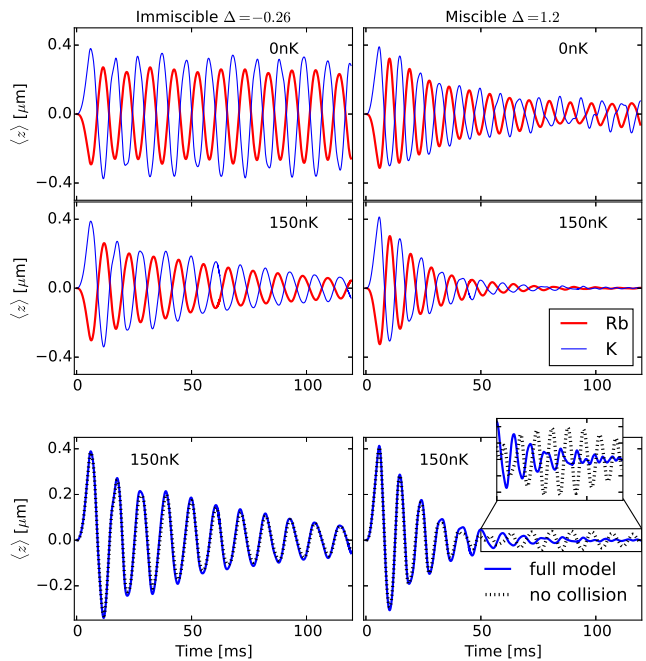}
  \vspace*{-5mm}
  \caption{\label{fig:aniso_dipole_osc} (Color online) The COM $\langle z\rangle$ of $^{87}$Rb (thick red) and $^{39}$K (thin blue) condensates at temperature $T= 0\,$nK (top) and $150\,$nK (middle) for different miscibilities. Bottom panels reveal the $^{39}$K COM with our full model (solid blue line), or the reduced version excluding all collisions (dashed black line)
The simulation parameters are the same as Fig.~\ref{fig:aniso_frac_den}.}
\end{figure}

In Fig.~\ref{fig:aniso_dipole_osc}, we show the simulated displacements of the condensates. The top panels display the undamped oscillation at zero temperature for an immiscible mixture (left) and a rapidly damped oscillation for a miscible mixture (right). At finite temperature (middle panels), numerical solution of our full binary kinetic model~\cite{edmonds_lee_2015a,*edmonds_lee_2015b} (including all possible collisional processes), reveals damping even for the immiscible mixtures, due to the 
 interaction with the thermal cloud. These numerical results are compared for different conditions by fitting the COM of the $^{39}$K atoms with a damped sinusoidal function for time $t>10\,$ms~\footnote{For temperature $T=0\,$nK, we restrict our fitting domain to $10\,$ms$<t<$60$\,$ms.}, as $\langle z\rangle = A\exp(-\gamma t)\cos(2\pi\nu t + \varphi)$. Both the damping rate $\gamma$ and the frequency $\nu$, shown in Figs.~\ref{fig:aniso_den_damping}(b) and \ref{fig:aniso_den_damping}(c) respectively, increase markedly at $\Delta\approx0.2$ (vertical dashes), at which $\Delta n_{\rm norm}$ also starts to change in Fig.~\ref{fig:aniso_den_damping}(a). We have checked the results for other atom numbers and arrived at the same conclusion (e.g. critical $\Delta\approx0.5$ if $N_{\rm Rb}=5\times10^4$ and $N_{\rm K}=10^5$, or critical $\Delta\approx0$ if $N_{\rm Rb}=5\times10^5$ and $N_{\rm K}=3\times10^4$). 

Interestingly, we find that the mean-field interactions between the BECs and the thermal clouds are by far the dominant damping mechanism; nonetheless, inclusion of collisions appears essential at long timescales ($\gg \gamma^{-1}$) to eliminate residual centre of mass oscillations, as evident by the $^{39}$K COM oscillations shown in the bottom panels of Fig.~\ref{fig:aniso_dipole_osc} (and associated inset).

It is noteworthy that the sharp changes in both quantities remain detectable at finite temperature (green squares and red diamonds in Fig.~\ref{fig:aniso_den_damping}), albeit with a decrease in the difference between the damping rates across the transition. This indicates the relevance of these changes in a realistic experimental setup. In addition, the much lower oscillation frequency compared to any of the axial trap frequencies (166$\,$Hz and 248$\,$Hz) highlights the impact of interspecies repulsion~\footnote{A detailed study of the frequencies to be published elsewhere.}.

\begin{figure}[!t]
  \includegraphics[width=0.49\textwidth]{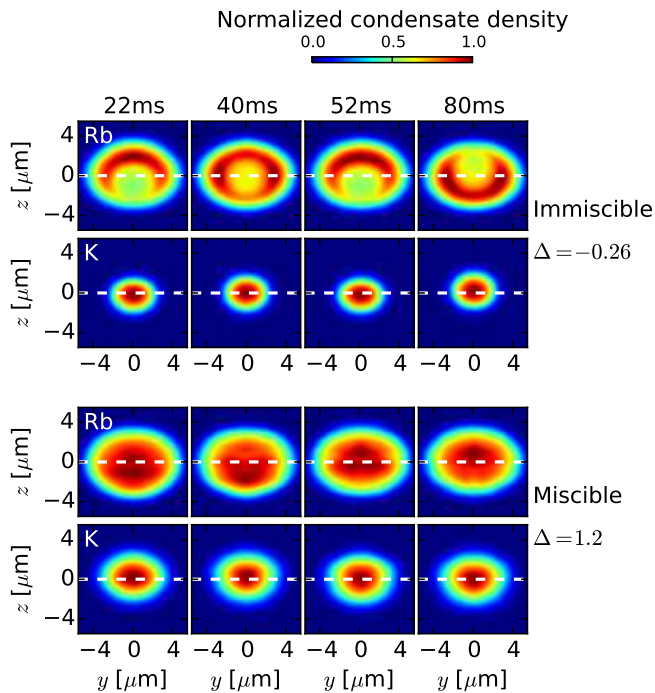}
  \vspace*{-5mm}
  \caption{\label{fig:aniso_dipole_snapshot} (Color online) Snapshots of the column-integrated density profiles of $^{87}$Rb and $^{39}$K condensates at $T=0\,$nK of an immiscible (top) and miscible (bottom) mixtures, showing the different dynamical excitations when the clouds execute dipole oscillation. The white dashes mark the initial trap center. The simulation parameters are the same as Fig.~\ref{fig:aniso_frac_den}. See online supplementary material for videos.}
\end{figure}

Overall, our results in Fig.~\ref{fig:aniso_den_damping} demonstrate that $\Delta n_{\rm norm}$ is physically-meaningful, and we can detect the transition by monitoring both the damping rates and the frequency. Interestingly, $\Delta n_{\rm norm}$ also reflects the penetrability of the mean-field barrier. Consequently, we can reinterpret our results as related to the distinct dynamical behavior of the BEC in the presence of a penetrable or impenetrable barrier, as seen also in recent experiments on vortex generation~\cite{kwon_seo_2015} and Josephson junction~\cite{valtolina_burchianti_2015} in single-component Bose gas.

It is interesting to note how counter-intuitive the underdamped oscillation of an immiscible mixture is. Since an immiscible mixture is associated with strong interspecies repulsion, we would naively expect the dipole oscillation to be quickly damped by the strong interaction. Our numerical results reveal the opposite.

This can be explained based on the different dynamical excitations associated with the topology of the equilibrium densities. In the immiscible case, there is a large interfacial tension between the two condensates~\cite{deng_schaeybroeck_2016}. Since a large amount of energy is needed to break the surface tension, the lower energy excitation involves undamped surface shape oscillations~\cite{svidzinsky_chui_2003}. In the top panels of Fig.~\ref{fig:aniso_dipole_snapshot}, we show the column-integrated density of the $^{87}$Rb condensate, $n_{\rm int}(y,z) = \int dx\;n_{c,1}\left(\sqrt{x^2+y^2},z\right)$, at different times. These snapshots clearly reveal the surface oscillation, where the $^{87}$Rb cloud encircles the $^{39}$K cloud and periodically displays a gap at the interface. 

For the miscible case, large spatial overlap of the condensates favour the counterflow instability~\cite{takeuchi_ishino_2010,suzuki_takeuchi_2010}. Sound, seen as density modulation in the bottom panels of Fig.~\ref{fig:aniso_dipole_snapshot}, can be generated if the relative speed $v$ of the two condensates is greater than $c_1+c_2$~\cite{castin_ferrier-barbut_2015,*abad_recati_2015}, where $c_i=\sqrt{g_{ii}n_{c,i}(0,0)/m_i}$ is the speed of sound of each individual condensate. This has been probed experimentally with a double superfluid Bose-Fermi mixture executing dipole oscillations in an elongated trap~\cite{ferrier-barbut_delehaye_2014,delehaye_laurent_2015} via a different excitation scheme. For our parameters, $v\approx (\omega_{z,1}+\omega_{z,2})\times 0.1\,\mu{\rm m}\approx0.3\,$mm/s, damping emerges due to density inhomogeneities even though $v$ is much smaller than the trap-center speeds of sound, $c_1\approx2.2\,$mm/s and $c_2\approx3.4\,$mm/s. In addition, the increase in damping rate with decreasing $\Delta$ for $\Delta > 0.5$ [Fig.~\ref{fig:aniso_den_damping}(b)] is consistent with the counterflow instability~\cite{chevy_2015}.



\emph{Experimental feasibility.} The simulations presented so far apply to many existing multi-species experiments, e.g. the mixture of $^{87}$Rb and $^{39}$K~\cite{wacker_jorgensen_2015}. Typically however, the gravitational sag between the two components of different mass is not compensated and experiments are conducted with relatively small spatial overlap~\cite{modugno_modugno_2002, pasquiou_bayerle_2013, wacker_jorgensen_2015, wang_li_2016}. Depending on the trap parameters the differential sag can exceed the typical Thomas-Fermi radius and therefore needs to be compensated to conduct the experiments proposed above.

The gravitational sag for each of the species is given by $g/\omega_{z,j}^2$, where $g$ is the acceleration due to gravity. The differential sag between both species can be canceled by employing an optical potential with a carefully selected wavelength. One option is to select a dipole trap which leads to identical frequencies $\omega_{z,j}$ for both species~\cite{ospelkaus_ospelkaus_2008}. Another option is to add an additional optical potential to an existing trap which introduces an upwards force proportional to the atomic mass, thus canceling the sag.

In both cases the necessary detuning $\delta_{j}$ of the dipole potential can be found as follows. In the simplified case of a two-level system the dipole force is proportional to $\Gamma_{j} /\nu_{0,j}^3 \delta_{j}$ where $\Gamma_{j}$ is the linewidth and $\nu_{0,j}$ is the transition frequency~\cite{grimm_weidemuller_2000}. Hence the same upwards acceleration for both species can be obtained if the criterion $m_2/m_1 = \Gamma_2 \nu_{0,1}^3 \delta_1 /  \Gamma_1 \nu_{0,2}^3 \delta_2$ is met. The detunings $\delta_{j}$ then provide the wavelength necessary to cancel gravity. A full calculation for the $^{87}$Rb-$^{39}$K mixture shows that an additional beam at 806~nm should hence be used to realize the second cancellation method outlined above. 

\emph{Conclusions.} We have provided numerical evidence that, for a trapped condensate mixture with overlapping trap centres, the miscible-immiscible transition depends critically on the condensate numbers, deviating from the simple homogeneous prediction used to date. We demonstrate that this transition can be mapped out experimentally by measuring the damping rate and the frequency of the dipole oscillations, the predominant contribution to which stems from mean-field coupling. We relate this change in damping rate across the transition to the different dynamical excitations due to the topology of the initial density distribution. The successful implementation of the fully self-consistent dynamical kinetic model for binary mixtures opens up a multitude of possibilities for studying coupled binary dynamics, and we hope that our work will inspire future experiments on systematic studies of the dynamical behaviour of trapped multicomponent condensates.

Data supporting this publication is openly available under an `Open Data Commons Open Database License'~\footnote{Additional metadata are available at: http://dx.doi.org/10.17634/122626-3.  Please contact Newcastle Research Data Service at rdm@ncl.ac.uk for access instructions.}. 

\emph{Acknowledgments.} KLL and NPP acknowledge discussion with F. Chevy and C. Salomon, and support from EPSRC Grant No. EP/K03250X/1 (KLL, NPP). NBJ, LW and JJA thank the Danish Council for Independent Research, and the Lundbeck Foundation for support. IKL was supported by the Ministry of Science and Technology, Taiwan (Grant No. MOST-103-2917-I-018-001). This work made use of the facilities of N8 HPC Centre of Excellence, provided and funded by the N8 consortium and EPSRC (Grant No.EP/K000225/1). The Centre is co-ordinated by the Universities of Leeds and Manchester. 

\bibliography{binaryPhaseDipole}

\newpage

\setcounter{figure}{0}

\appendix*
\section{Full phase diagrams}
\begin{figure}[!b]
  \includegraphics[width=0.49\textwidth]{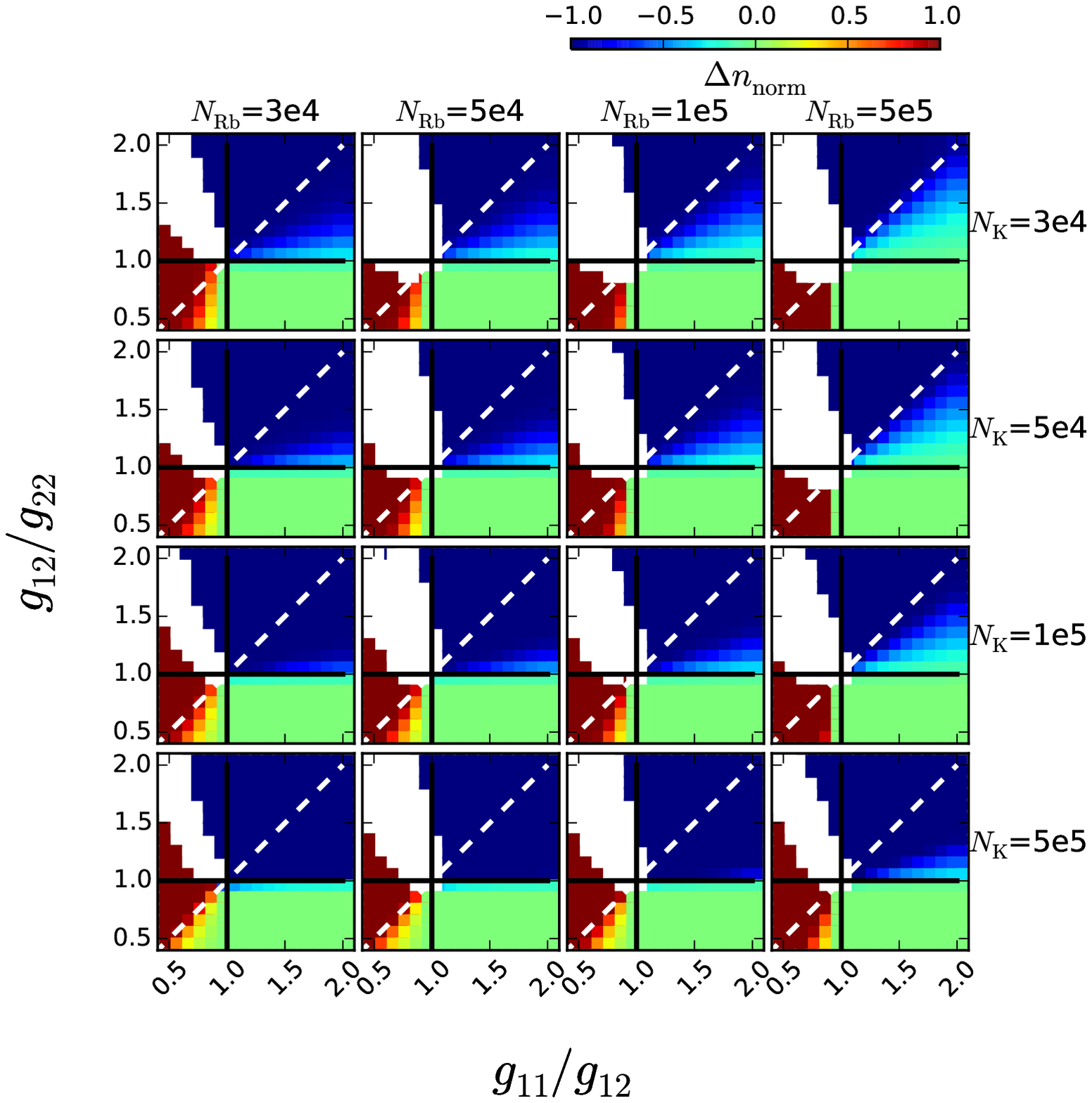}
  \caption{\label{fig:aniso_phaseT0} (Color online) Zero-temperature phase diagram of $^{87}$Rb-$^{39}$K mixtures with different total number of atoms ($N_{\rm Rb}$, $N_{\rm K}$) in anisotropic harmonic traps.   The trap frequencies are $\nu_r = 119\,$Hz (178$\,$Hz) and $\nu_z = 166\,$Hz (248$\,$Hz) for $^{87}$Rb ($^{39}$K) atoms. $N_{\rm Rb}$ increases from left to right while $N_{\rm K}$ increases from top to bottom. The symmetrical phase is characterised by a normalised trap-center density $\Delta n_{\rm norm}$ [see Eq.~\eqref{eq:n_norm}] while the asymmetrical phase is shown in white.}
\end{figure}

Here, we demonstrate the generality of the presented resuts, by considering all 16 possibilities for the stability phase diagram based on experimentally-realistic numbers of $^{87}$Rb and $^{39}$K (see Fig.~\ref{fig:aniso_phaseT0}).

\begin{figure}[!t]
  \includegraphics[width=0.49\textwidth]{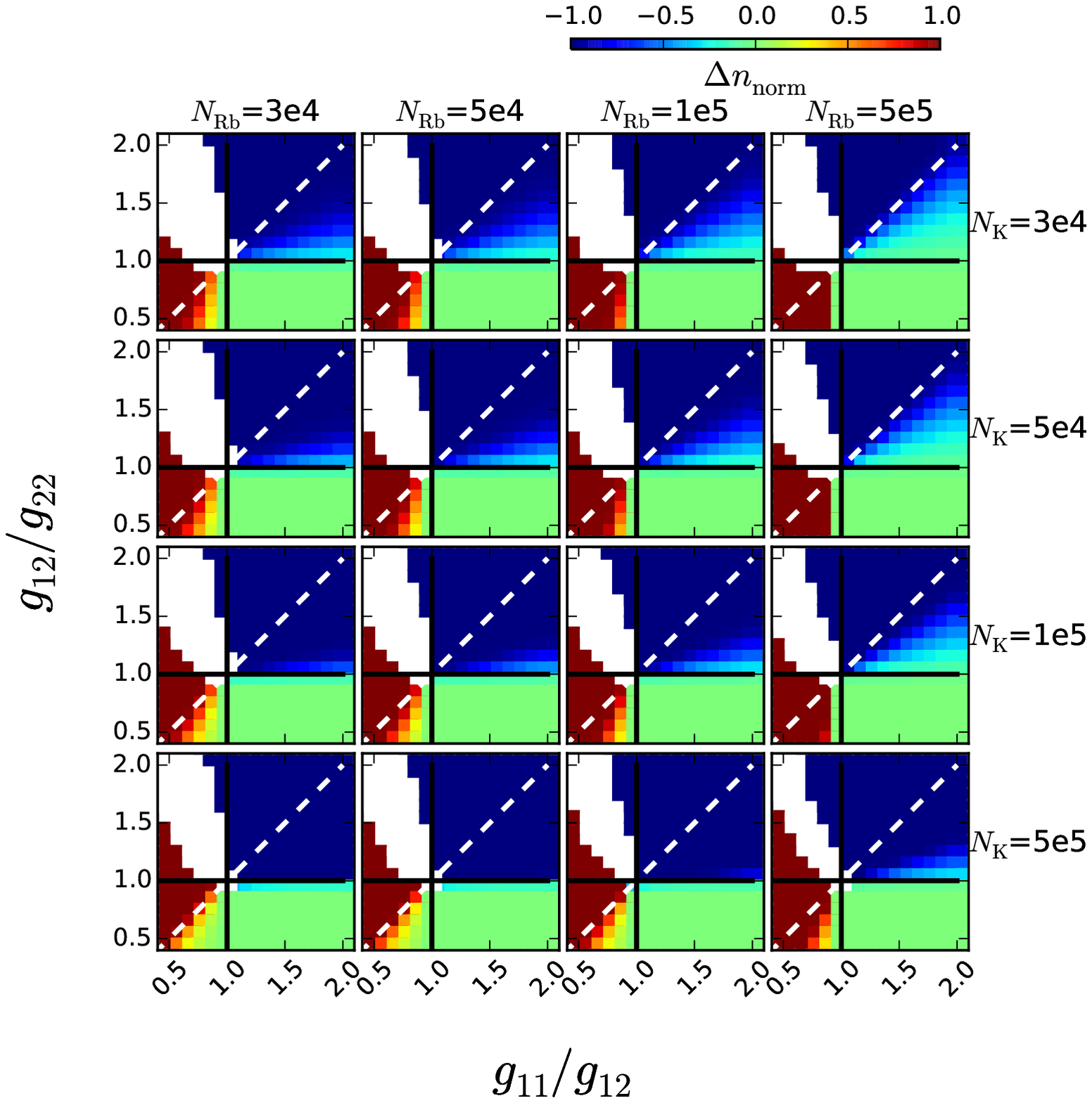}
  \caption{\label{fig:iso_phaseT0} (Color online) Same as Fig.~\ref{fig:aniso_phaseT0} but for atoms in isotropic harmonic trap. The trap frequencies are $\nu = 133\,$Hz (199$\,$Hz) for $^{87}$Rb ($^{39}$K) atoms.}
\end{figure}

To rule out trap anisotropy as the underlying cause of our observations, we also perform identical simulations but for atoms in isotropic trap (see Fig.~\ref{fig:iso_phaseT0}). Similar features that have been discussed in the text are clearly visible.

Our results thus confirm that, in general, and for a broad range of realistic parameters, the transition from miscible to immiscible does not happen precisely at the assumed $\Delta=0$ line.

\end{document}